\documentclass{article}
\usepackage{spconf,graphicx,amsmath,amsfonts,amssymb,bm, soul}
\usepackage[colorlinks=true, linkcolor=blue, citecolor=blue, urlcolor=blue]{hyperref}
\hypersetup{pdfcreator  = {LaTeX with hyperref package},
               pdfproducer = {dvips + ps2pdf} }

\title{Cylindrical coordinates for LiDAR Point cloud compression }
\name{Shashank N. Sridhara,  Eduardo Pavez,  Antonio Ortega}
\address{University of Southern California, Los Angeles, California, USA}
\begin{document}
\ninept
\maketitle
\begin{abstract}
We present an efficient voxelization method
to encode the geometry and attributes of 3D point clouds obtained from autonomous vehicles. 
Due to the circular scanning trajectory of sensors,
the geometry of LiDAR point clouds is inherently different from that of point clouds captured from RGBD cameras. Our method exploits these specific properties to representing points in \textit{cylindrical coordinates} instead of conventional Cartesian coordinates. We demonstrate that Region Adaptive Hierarchical Transform (RAHT) can be extended to this setting, leading to attribute encoding based on a volumetric partition in cylindrical coordinates. Experimental results show that our proposed voxelization outperforms conventional approaches based on Cartesian coordinates for this type of data. We observe a significant improvement in attribute coding performance with 5-10\% reduction in bitrate and octree representation with 35-45\% reduction in bits.
%
%
\end{abstract}
\section{Introduction}
\label{sec:intro}
Advancements in sensor technology coupled with their wide applications in autonomous driving, augmented and virtual reality (AR/VR) have made point clouds ubiquitous in 3D scene understanding. While RGBD cameras are used to capture  point clouds in AR/VR applications, light detection and ranging (LiDAR) sensors are predominantly used in remote sensing and autonomous driving. LiDAR sensors are extensively used in autonomous driving because they  provide more precise 3D scene representations as compared to depth estimation based on 2D images. Since every autonomous vehicle collects real time LiDAR sweeps, the amount of data generated is immense. As an example, a single LiDAR sweep of a Velodyne HDL-64 LiDAR generates over 100,000 points, resulting in approximately 80 billion points per day \cite{octsqueeze}. Several large scale LiDAR point cloud datasets, specifically for autonomous driving research, are publicly available \cite{kitti,waymo, nuscenes, pandaset}. 

A point cloud consists of a list of 3D point coordinates along with the corresponding attributes. A real time LiDAR sweep from an autonomous vehicle generates as an attribute the intensity of the returning (reflected) beam value, $a_{i} \in {\rm I\!R}$. Compression of both geometry and attributes is very important because the point cloud sequence has to be stored and transmitted efficiently. Geometry based point cloud compression (G-PCC) is commonly used to compress attributes of various types of point clouds, including LiDAR point clouds \cite{mpeg}. G-PCC methods exploit the geometry information to encode the associated attributes. Region Adaptive Hierarchical Transform (RAHT) \cite{raht}, a multi-resolution orthonormal transform is the current state-of-the-art attribute compression method for LiDAR point clouds. Since the RAHT encoder is computationally efficient and robust to all types of point clouds, it has become a part of the  MPEG point cloud compression standard for attribute coding \cite{mpeg}. 
%

To apply G-PCC to a dataset, the first step is \textit{voxelization}, where points are assigned to voxels, which play the same role as pixels for images. %
Conventionally, voxelization is performed using cubes at multiple resolutions, so that the volume of the smallest (voxel) cube determines the point cloud resolution. Typically, the geometry of a cubic voxelized 3D point cloud is represented as an {\em octree} \cite{octree}. The octree data structure can be efficiently traversed and lends itself to transform techniques for compression  \cite{raht,ragft,block-gft}.
\begin{figure*}[htb]
\centerline{\includegraphics[width=0.80\textwidth]{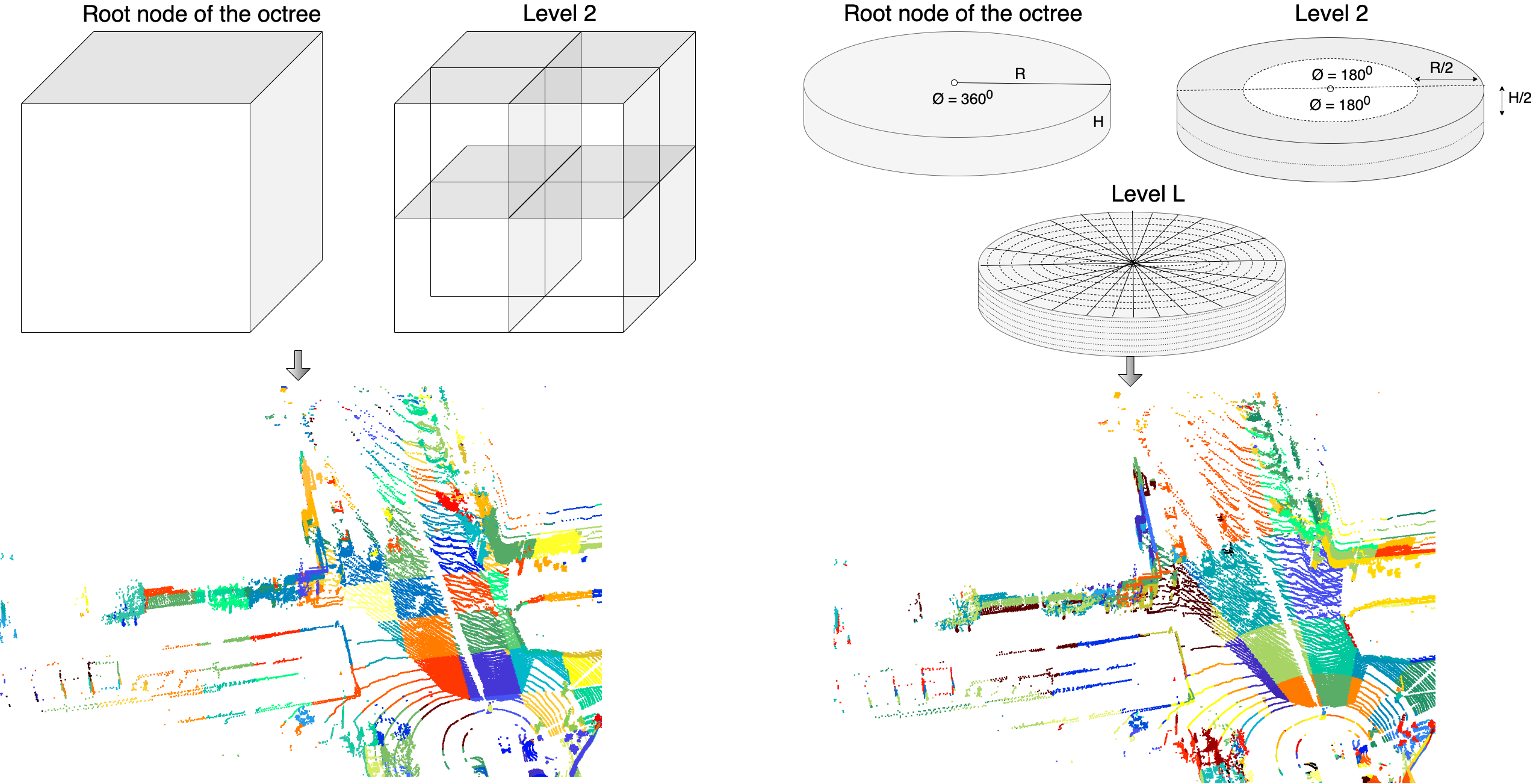}}
\caption{Comparison between voxelization in Cartesian and cylindrical coordinates. Uniform partitioning of the space into cubes (left). Cylindrical volume elements partition the space non-uniformly (right).  The volume of cylindrical voxels increases as we move away from the sensor which will allow us to capture more number of points, thereby avoiding uneven point distribution across the cells.}
\label{fig:cartesian_vs_cylindrical}
\end{figure*}

The main motivation for our work comes from the observation that LiDAR point clouds from autonomous vehicles are structurally different from those used in AR/VR applications.  
%
First, LiDAR point clouds exhibit circular patterns due to the scanning trajectory of the sensor. 
Moreover, since beams are sent out from a sensor in the center of the observed region, spacing between neighboring beams increases away from the sensor, so that point density decreases away from the origin.
Thus,   
existing compression methods that represent the point cloud as a cubical volume \cite{octsqueeze} 
will result in grid cells near the sensor (origin) having more points, while  peripheral grid cells may contain very few points. This uneven partitioning of the points is not only computationally inefficient, but also fails to take into account the inherent geometry of the point cloud. 


In this paper, 
we use cylindrical voxelization, with the goal of better leveraging the circular geometry of LiDAR point clouds. By using  cylindrical coordinates to partition the space, voxels that are away from the sensor have larger volume compared to that of those near the sensor. This non-uniform partitioning is  suitable for LiDAR point clouds, which have higher point density near the origin and very few sparse points away from the sensor. Moreover, with a cylindircal voxelization we can still build an octree structure, so that existing methods for point cloud processing and G-PCC can be applied with relatively small changes. 
This proposed voxelization strategy is shown to
reduce significantly octree size, 
while resulting in improved attribute compression using RAHT.
%
Non-Cartesian coordinate systems to represent LiDAR point clouds have  become popular in 3D computer vision applications such as object detection and semantic segmentation \cite{polarnet, spherical_kernel,cylindrical}. However, to the best of our knowledge,   alternate coordinate systems, e.g., cylindrical/spherical, have not been used for compression of LiDAR point clouds generated from autonomous vehicles.

The rest of the paper is organized as follows. Section \ref{sec:2}   describes voxelization  in cylindrical coordinates, while Section \ref{sec:raht_cyl} presents  attribute coding using RAHT in cylindrical coordinates. Experiments and conclusions are in Sections \ref{sec:experiments} and \ref{sec:conclusion} respectively.

\section{Voxelization}
\label{sec:2}

\subsection{Voxelization in cylindrical coordinates}
Consider a point cloud from a single LiDAR sweep, containing $N$ points along with their  attributes. 
Each point is represented as $P_{i} = [x_{i}, y_{i}, z_{i}, a_{i}]: i= [1,....,N]$ where $(x_{i}, y_{i}, z_{i})$ is the location of the point in Cartesian coordinates and $ a_{i}$ is the attribute (intensity)  of the point. 
In the \textit{Cartesian} coordinate system, voxelization partitions the space into uniform volume elements (cubes) called voxels. All points inside a voxel are quantized to the voxel center and their attributes are averaged. 
 After voxelization, the geometry of the point cloud is encoded using octree scanning. If the volume of the cube that represents the entire space is $W \times W \times W$, then the first level of octree partitions the cube into $2^{3}$ cubes, each with volume ${W}/{2} \times {W}/{2} \times {W}/{2}$. After $L$ levels of partitioning there will be $2^{3L}$ cubes, each with volume ${W}/{2^{L}} \times {W}/{2^{L}} \times {W}/{2^{L}}$. 
The choice of $L$ determines the maximum point cloud resolution. 
At any level, the octree includes information about the occupancy of each cube, and only those cubes containing points are further partitioned at the next level.
Thus, in an octree with $L$ levels, the root node corresponds to a cube enclosing all points, whereas the leaves of the octree represent the set of occupied voxels. 
 
Our proposed voxelization in cylindrical coordinates is analogous, leading to a partition of the space into cylindrical volume elements. The points representing the location in Cartesian coordinates $(x, y, z)$ are converted to cylindrical coordinates $(r, \theta, h)$, where
$
    r = \sqrt{x^{2} + y^{2}}, \quad 
    \theta = \tan^{-1} \left({y}/{x}\right), \textnormal{ and} \quad 
    h = z.
$
Assume all points are inside a cylinder of volume $V = \pi R^{2} H $.
We consider a general separable partitioning strategy, with $J$ radial divisions, $K$ angular divisions and $L$ resolutions in height. While in the Cartesian coordinate case, all sides of a cube were divided by two as resolution increases, here we allow different scaling 
along each direction, as described below.   
The resulting octree in cylindrical coordinates results in $2^{J+K+L}$ cylindrical volume elements, each with volume $dV = r\cdot dr  \cdot d\theta \cdot dh$. For simplicity, 
in this paper we divide all the dimensions to the same depth i.e., $J = K =L$. Figure \ref{fig:cartesian_vs_cylindrical} illustrates the difference between Cartesian and cylindrical voxelization.

\begin{figure}[t]
\centerline{\includegraphics[width= 0.5\textwidth]{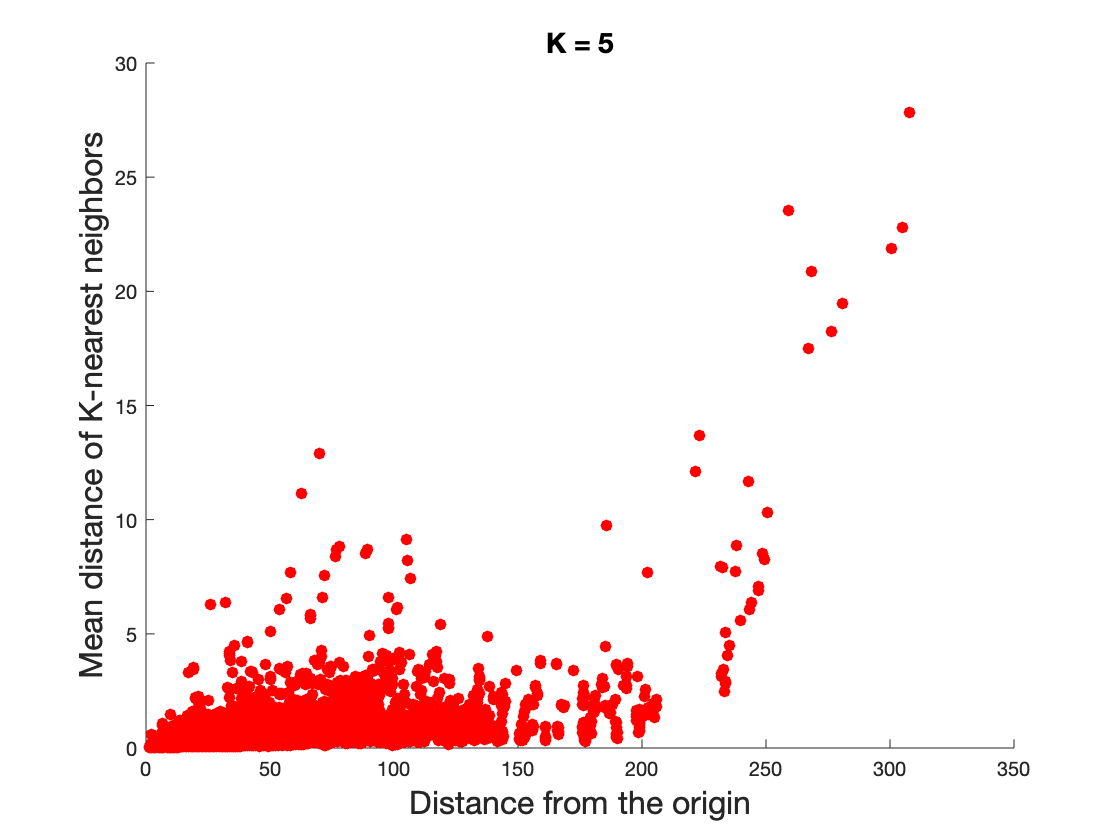}}
\caption{Scatter plot of mean distance of $K$ nearest neighbors ($K = 5$) as the function of  the radial distance from the origin. It can be seen that points far away the origin are also far away from their $K$ nearest neighbors, which corresponds to a decrease in point density. }
\label{fig: log_res}
\end{figure}

The cylindrical voxelization naturally adjusts the size of voxels based on the distance from the sensor. Since the point density rapidly decreases as we move away from the sensor, partitioning the radial dimension uniformly 
will result in voxels with high variations in  point density as shown in Figure \ref{fig: log_res}, where average distance between points increases significantly as the radial distance increase. To address this issue, we propose to  partition the radial dimension logarithmically ($log(r)$) resulting in smaller volume  partitions near the sensor, where point density is high, and larger volume ones far away from the sensor,  where point density is lower. 
%


\begin{table}[]
\begin{tabular}{|l|c|c|}
\hline
 &
  \begin{tabular}[c]{@{}c@{}}number of volume\\  elements\end{tabular} &
  \begin{tabular}[c]{@{}c@{}}avg. number of points in \\ each volume element\end{tabular} \\ \hline
cartesian &
  3925 &
  44.35 \\ \hline
cylindrical &
  1334 &
  130.50 \\ \hline
\end{tabular}
\caption{Number of volume elements and average number of points in each volume element when the point cloud is voxelized in Cartesian and cylindrical coordinates for 8 levels. These results are obtained for one of the point clouds in \cite{pandaset}.}
\label{tab:stats}
\end{table}
Results in Table \ref{tab:stats} show that  peripheral voxels have larger volume compared to that of the voxels near the sensor. Thus, with cylindrical voxelization the point cloud can be represented with fewer voxels and
this
leads to a more balanced partitioning, where voxels tend to have a more uniform number of points. As a consequence, the number of bytes required to store the octree will be reduced significantly. The results are discussed in detail in Section \ref{sec:experiments}.
\begin{figure}[t]
\centerline{\includegraphics[width=0.5\textwidth]{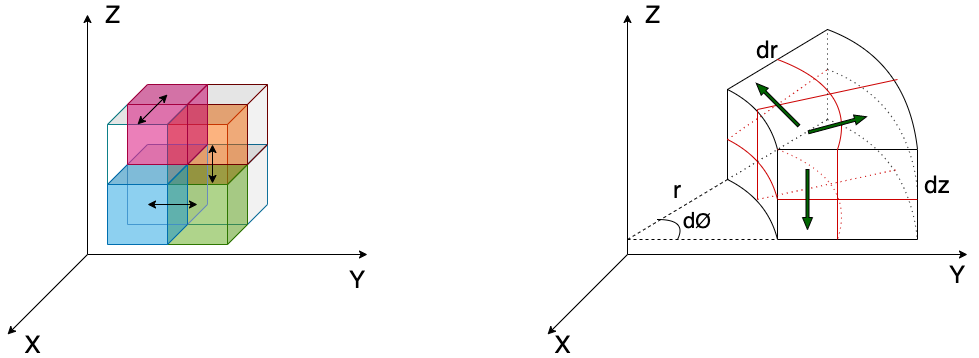}}
\caption{One level of RAHT applied to a cube of 8 voxels (left). One level of RAHT applied to a cylindrical volume of 8 voxels (right)}
\label{fig: raht_cyl}
\end{figure}


\subsection{Voxelization error in Cartesian and cylindrical coordinates}
Consider a point cloud contained in a cube of volume $ W \times W \times W $ to  be voxelized to a depth of $J$. Let $x,y,z$ be the Cartesian coordinates of a point,  and let $\hat{x}, \hat{y},\hat{z}$ be the voxelized coordinates, that is
\begin{equation*}
    \hat{x} = Q\cdot round\left(\frac{x}{Q}\right),
    \hat{y} = Q\cdot round\left(\frac{y}{Q}\right),
    \hat{z} = Q\cdot round\left(\frac{z}{Q}\right),
\end{equation*}
where $Q = W/2^{J}$. The voxelization error of each coordinate is denoted by $\hat{x} = x + \epsilon_1$, $\hat{y} = y + \epsilon_2$ and $\hat{z} = z + \epsilon_3$, thus we have
\begin{equation}
    E_{cart} = (x - \hat{x})^2 +(y - \hat{y})^2 + (z - \hat{z})^2  = \epsilon_1^2 + \epsilon_2^2 + \epsilon_3^2.
\end{equation}
Assuming that the error variables $\epsilon_i$ are independent,  zero mean with variance $\sigma^2$, the expected error is equal to $\mathbb{E}[E_{cart}] = 3\sigma^2$. This indicates that all axis have the same influence over the total error. 

However, when we voxelize the point cloud in cylindrical coordinates, we observe a different phenomenon. The cylindrical coordinates   $(r,\theta, h)$ are discretized as in the Cartesian case, but with different quantization parameter, thus 
$\hat{r} = Q_1\cdot round\left({r}/{Q_1}\right),
\hat{\theta} = Q_2\cdot round\left({\theta}/{Q_2}\right),
\hat{h} = Q_3\cdot round\left({h}/{Q_3}\right)$. The quantization  parameters are $Q_1 = W_{R}/2^{J} $, $Q_2 = W_{\theta}/2^{J}$, and $Q_3 = W_{H}/2^{J}$.
Then the reconstruction formula from cylindrical to cartesian coordinates is $\hat{x}=\hat{r}cos(\hat{\theta})$, $\hat{y} = \hat{r} sin(\hat{\theta})$, and $\hat{z} = \hat{h}$.
The voxeliation error for each coordinate is $\hat{r} = r + e_1$, $\hat{\theta} = \theta + e_2$ and $\hat{h} = h + e_3$ and 
the total voxelization error in cyclindrical coordinates $E_{cyl} = (x - \hat{x})^2 +(y - \hat{y})^2 + (z - \hat{z})^2$ is 
\begin{align}
E_{cyl}  = e_1^2 + 2r(r + e_1)(1-cos(e_2)) + e_3^2.
\end{align}
Assuming that $e_i$ are independent zero mean variables with variances $\sigma_i^2$, and using the Taylor expansion of cosine $\cos(\theta) \approx 1 - \theta^2/2$, the expected error is approximately,  
\begin{align}
    \mathbb{E}[E_{cyl}] \approx \sigma_1 ^2 + r^2 \sigma_2^2 + \sigma_3^2
\end{align}

The first and third terms correspond to errors in radial and height directions, and resemble those of Cartesian coordinates, since they are independent of the magnitude of the signal along those directions. The second term contains interactions between angular and radial directions, and suggests that when  the point $(x,y,z)$ is close to the origin ($r$ is small), this term is negligible, and thus cylindrical voxelization incurs smaller errors. When the point is far away from the origin ( $r$ large), cyclindrical voxelization has larger error.

\begin{figure}[b]
\centerline{\includegraphics[width=0.4\textwidth]{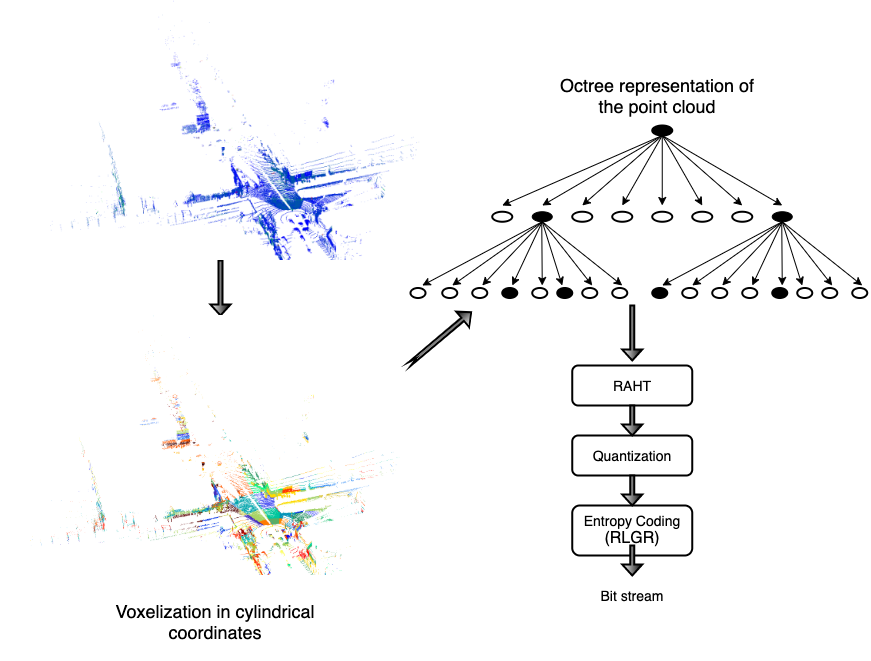}}
\caption{Encoding attributes of the LiDAR point cloud using RAHT by encoding the geometry using octree in cylindrical coordinates}
\label{fig: encoding_flow}
\end{figure}

\section{RAHT in cylindrical coordinates }
\label{sec:raht_cyl}
\begin{figure*}[t]
\centering
\includegraphics[width = 0.24\textwidth]{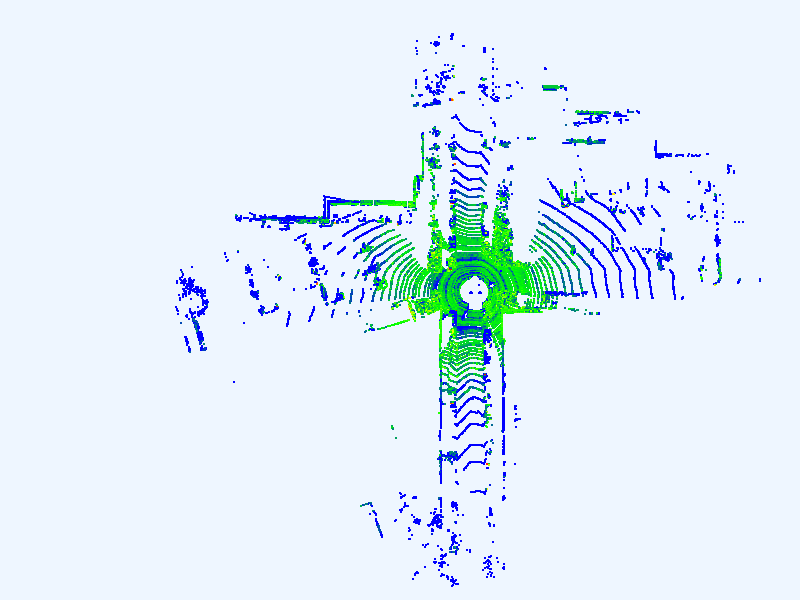}
\includegraphics[width = 0.24\textwidth]{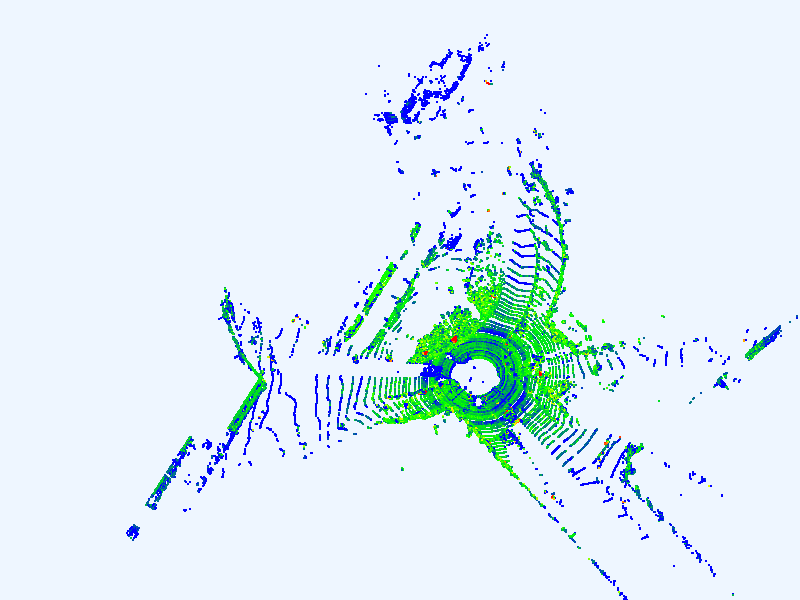}
\includegraphics[width = 0.24\textwidth]{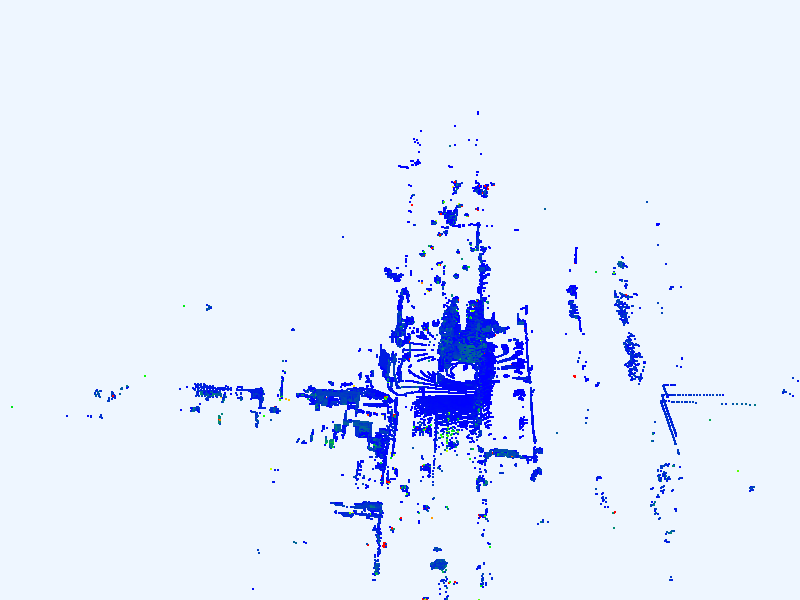}
\includegraphics[width = 0.24\textwidth]{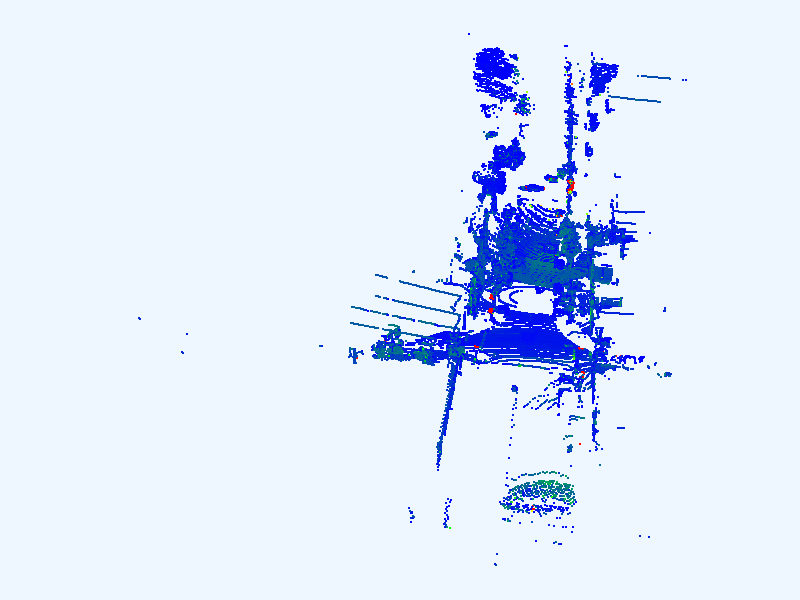} \\
\includegraphics[width = 0.24\textwidth]{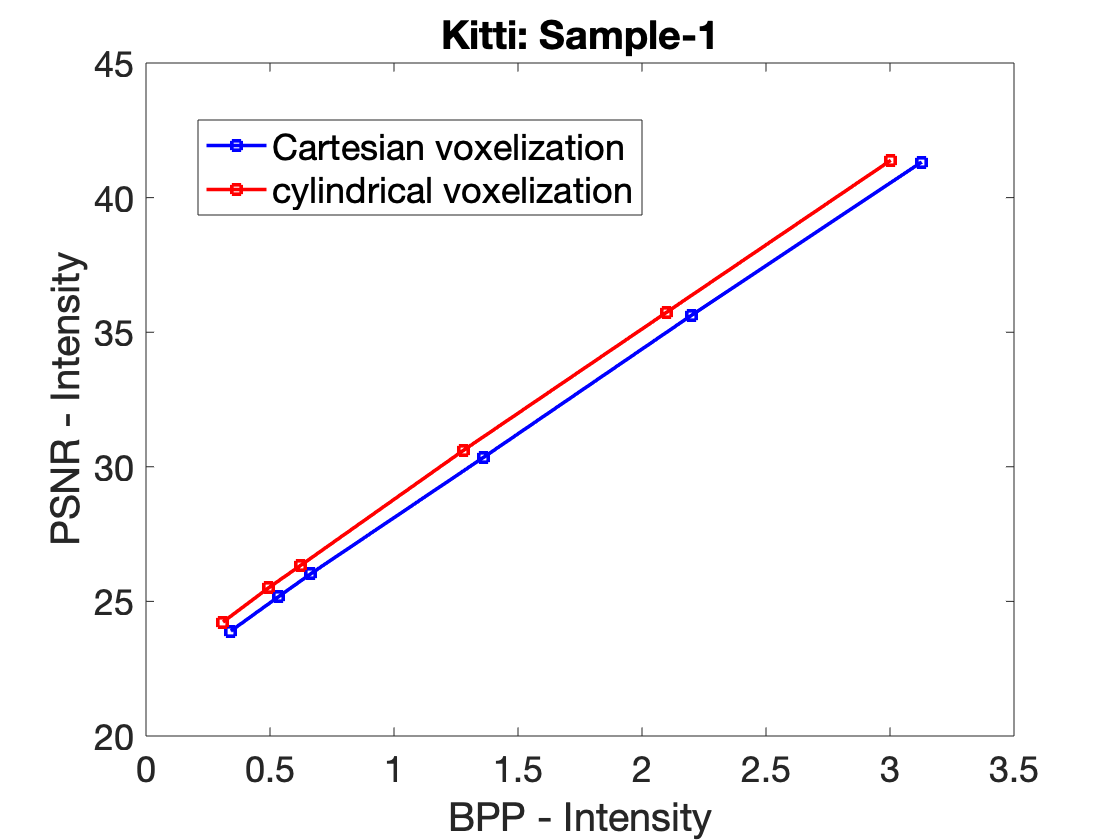}
\includegraphics[width = 0.24\textwidth]{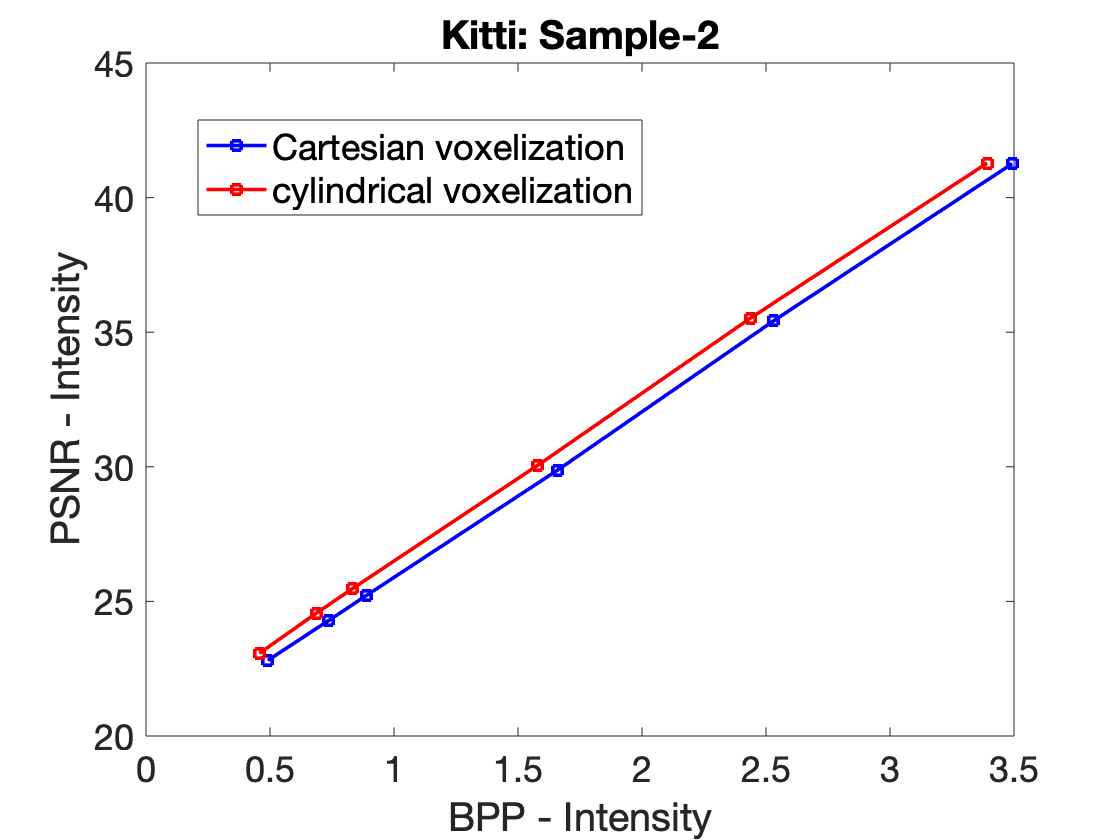}
\includegraphics[width = 0.24\textwidth]{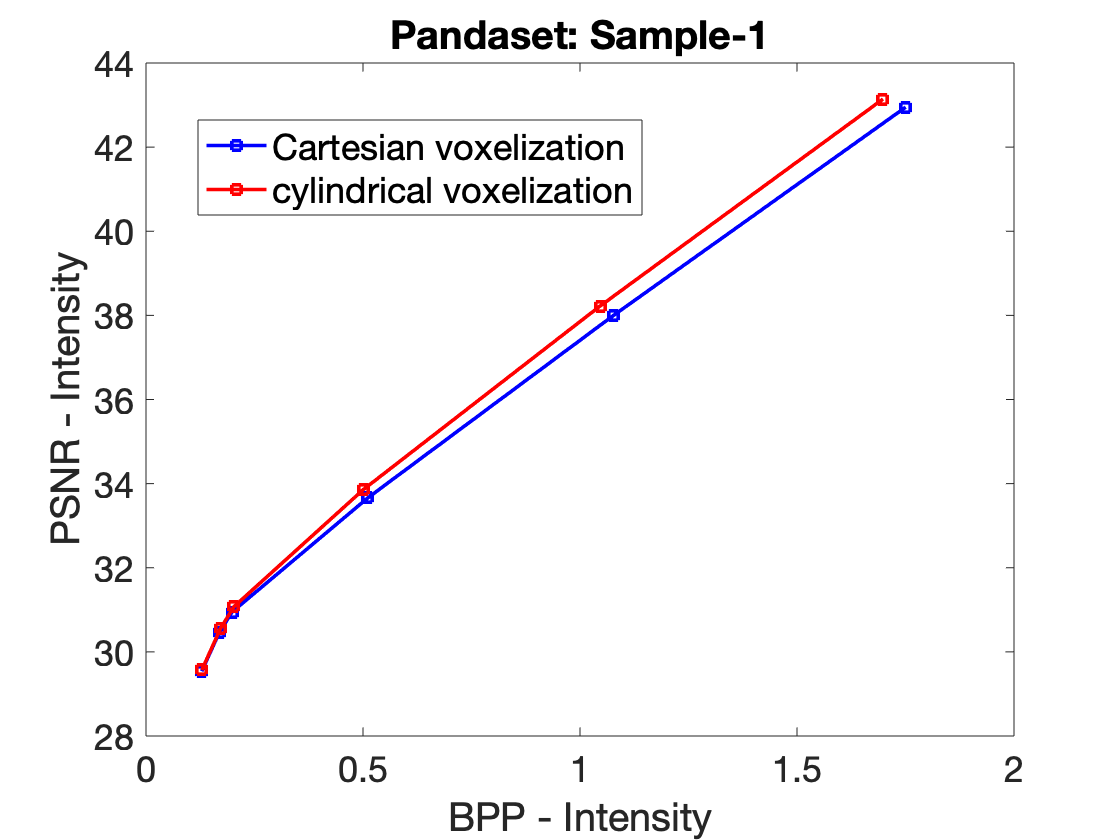}
\includegraphics[width = 0.24\textwidth]{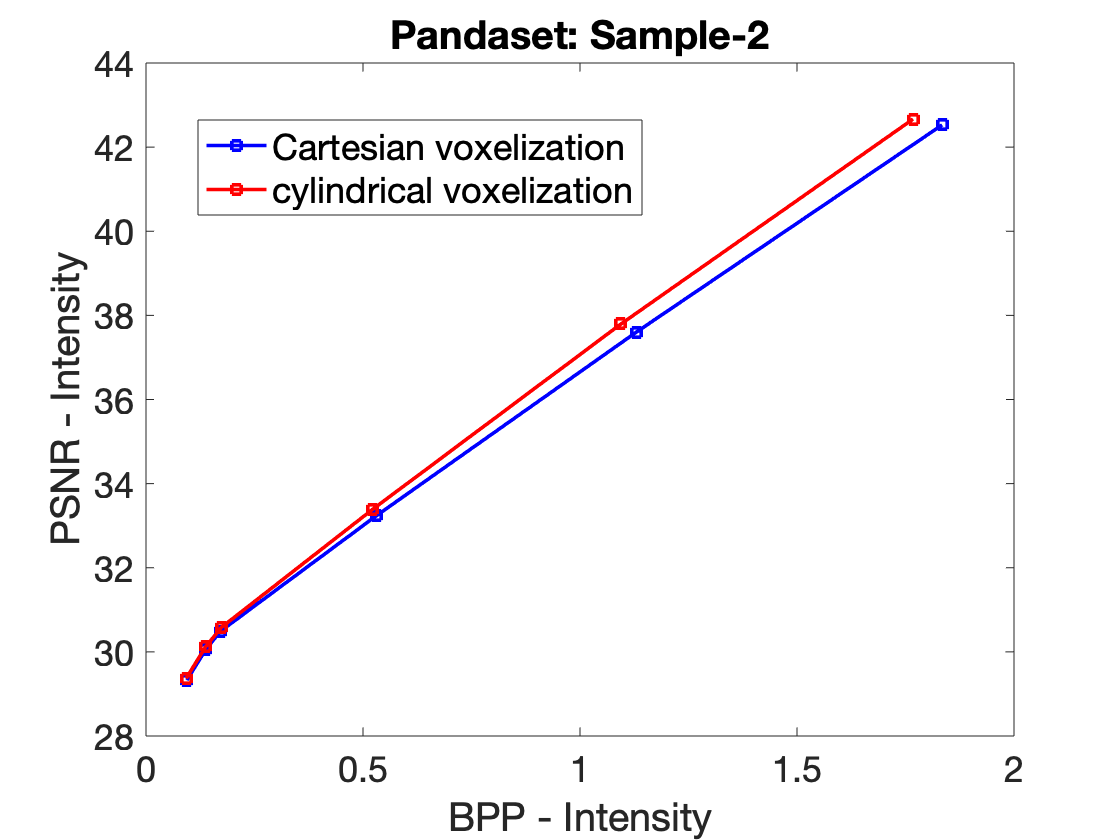}
\caption{Rate distortion curve for attribute compression on point clouds from kitti dataset and pandaset in Cartesian and cylindrical coordinates using RAHT}
\label{fig: rd_cruve}
\end{figure*}
Region adaptive hierarchical transform (RAHT) \cite{raht} is a multiresolution, orthonormal transform applied to the attributes of point clouds whose geometry is encoded using an octree. 
The main idea behind RAHT is to predict the attributes of nodes at the higher level of the octree (finer resolution)  using the attributes associated with the lower level nodes (coarser resolution). At each level, RAHT applies a forward transform by scanning the octree backwards and combining the voxels into larger ones until reaching the root node, which represent the entire space. The scan order is reversed while applying the inverse transform. See \cite{raht,dynamic_polygon_cloud} for implementation details.

Consider a point cloud whose geometry is encoded  using an octree with $L$ levels. A level $l < L$ in the octree represents a cube or a cylindrical volume element, depending on voxelization. When a forward transform of RAHT is applied at a particular level $l$, the algorithm picks a direction $(x,y,z)$ 
and checks for the occupied voxels to process along that direction. At every level, the forward transform outputs low pass and high pass coefficients. The high pass coefficients are preserved for entropy coding, while the low pass coefficients are propagated to the next level in the octree for further processing. After the tree is traversed backwards all the way to the root node, we will have one low pass coefficient and remainder high pass coefficients. The transform coefficients are uniformly quantized and entropy coded \cite{rlgr} to generate the bit stream.

The geometry of the point cloud is encoded using an octree and transmitted to the decoder separately. Although RAHT is applied on the attributes of the point cloud, the construction of the hierarchical decomposition mainly depends on the encoded geometry. As shown in Figure \ref{fig: raht_cyl}, for cylindrical coordinates, at each step recombining the voxels will be along $(r, \theta, h)$. Once the point cloud is represented in the form of an octree, RAHT can exploit the octree structure to compress the attributes. Therefore, we use RAHT for a space partitioned in any non-cartesian coordinate system by representing the geometry using an octree. Figure \ref{fig: encoding_flow} illustrates the entire flow of encoding the attributes.

%

%
\section{Experiments}
\label{sec:experiments}
We present our results on point clouds from Kitti dataset \cite{kitti} and PandaSet \cite{pandaset}. Both are large scale open source LiDAR datasets for autonomous driving which consists of point clouds from various driving scenes. We picked point clouds at different timestamps to evaluate the performance of both geometry and attribute compression in cylindrical coordinates. Initially, we compared the performance of RAHT on octrees with different number of levels constructed in Cartesian and cylindrical coordinates. RAHT achieved the best performance for $13$ and $16$ octree
levels in cylindrical and Cartesian coordinates, respectively, and these were the chosen octree depths in all our experiments. 
We use Bjontegaard metric \cite{bjontegaard} to compare the rate-distortion curves of attribute coding in Cartesian and cylindrical coordinates using RAHT. The Bjontegaard metric provides average PSNR difference and percentage bitrate saving  between those two rate-distortion curves. The distortion PSNR of the attribute (intensity) is given by,
\begin{equation}
     PSNR_{A} = -10  \log_{10} \left(\frac{{\Vert I - \hat{I}\Vert_{2}^{2}}}{{255^{2} N}}\right),
\end{equation}
%
where $I$ and $\hat{I}$ represent the original and decoded signals (attribute), and $N$ is the total number of points in the point cloud. The rate is reported in bits per point [bpp] $R = B / N$, where B is the number of bits used to encode the attributes. As shown in Figure \ref{fig: rd_cruve}, cylindrical voxelization achieved considerable gain in attribute coding over cartesian voxelization on both the datasets. The point clouds selected for our experiments represent different traffic scenarios and lighting conditions. While we report the results for 4 samples, we observed similar results across most point clouds. Table \ref{tab:attribute} shows average PSNR gain and percentage bitrate saving when cylindrical coordinates are used over Cartesian coordinates. 
\begin{table}[ht]
\centering
\begin{tabular}{|c|c|c|c|}
\hline
dataset & number of points & avg. PSNR gain   & bitrate saving                                   \\ \hline
kitti-1 & 120636 & 0.63 & 14.56 \% \\ \hline
kitti-2 & 123526 & 0.58 & 11.21 \% \\ \hline
pandaset-1 & 169729 &0.36 & 4.96 \% \\ \hline
pandaset-2 & 177951 &0.43 & 5.05 \% \\ \hline
\end{tabular}
\caption{Comparison of attribute compression with RAHT in Cartesian and cylindrical coordinates}
\label{tab:attribute}
\end{table}
As the geometry of the point cloud is encoded using an octree before compressing the attributes, we report the bitrate in bits per point [bpp] to store and transmit the octree in Cartesian and cylindrical coordinates. The bitrate to represent the geometry is significantly reduced when the octree is constructed in cylindrical coordinates. This is because the octree constructed in cylindrical coordinates is smaller ($13$ levels) when compared to the octree constructed in Cartesian coordinates ($16$ levels). Table \ref{tab:geometry} show the difference in bitrate to represent the octree in Cartesian and cylindrical coordinates. 
%
\begin{table}[ht]
\centering
\begin{tabular}{|c|c|c|c|}
\hline
dataset & cartesian octree & cylindrical octree & bitrate saving \\ \hline
kitti-1      & 40.24                 & 23.76                   &40.95\% \\ \hline
kitti-2      & 42.55                 & 23.23                   &45.40\%\\ \hline
pandaset-1      & 29.25                 & 20.00                   &31.62\%\\ \hline
pandaset-2      & 30.91                 & 19.99                   &35.32\%\\ \hline
\end{tabular}
\caption{Comparison of cartesian and cylindrical octrees (bpp).}
\label{tab:geometry}
\end{table}
\section{Conclusion}
\label{sec:conclusion}
In this paper, we use the notion of voxelization in cylindrical coordinates to compress the geometry and attributes of LiDAR point clouds from autonomous vehicles. This voxelization method can effectively exploit the circular geometry of the point cloud, as well as the varying point density. We demonstrate theoretically that cylindrical voxelization is more accurate near the origin, which is where LiDAR point clouds have higher point density. We show numerically that the proposed voxelization  reduces octree size, while also resulting in  attribute compression gains with RAHT. 



\label{sec:ref}

\bibliographystyle{IEEEbib}
\bibliography{strings,refs}

\end{document}